# Detecting Malicious Urls of COVID-19 Pandemic Using ML Techniques

Jamil Ispahany & Rafiqul Islam
Charles Sturt University, NSW, Australia.
{jamil.ispahany@csu.edu.au, mislam@csu.edu.au}

**Abstract:**

*Throughout the COVID-19 outbreak, malicious attacks have become more pervasive and damaging than ever. Malicious intruders have been responsible for most cybercrimes committed recently and are the cause for a growing number of cyber threats, including identity and IP thefts, financial crimes, and cyber-attacks to critical infrastructures. Machine learning (ML) has proven itself as a prominent field of study over the past decade by solving many highly complex and sophisticated real-world problems. This paper proposes an ML-based classification technique to detect the growing number of malicious URLs, due to the COVID-19 pandemic, which is currently considered a threat to IT users. We have used a large volume of Open Source data and pre-processed it using our developed tool to generate feature vectors and we trained the ML model using the apprehensive malicious threat weight. Our ML model has been tested, with and without entropy to forecast the threatening factors of COVID-19 URLs. The empirical evidence proves our methods to be a promising mechanism to mitigate COVID-19 related threats early in the attack lifecycle.*

## 1. Introduction

On the 31st of December 2019, unknown pneumonia was discovered in Wuhan, China. Since its detection, the outbreak of the novel coronavirus disease known as COVID-19 has spread throughout the world at an unprecedented rate, impacting the lives of almost every citizen across 213 countries and territories [1-2,13]. Whilst COVID-19 is gradually being controlled in many countries, it is still spreading throughout the world at a rapid pace. As of August 2020, COVID-19 had infected over 25 million people and responsible for the death of over 850 thousand people worldwide. Figure 1 shows the countries case distribution [1].

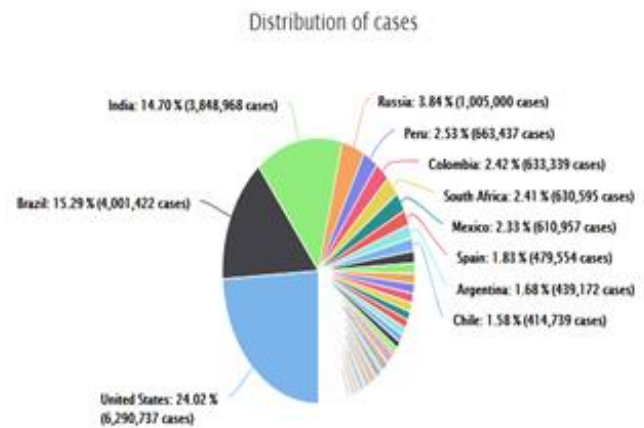

Fig. 1: Countries case distribution [1]

Due to the coronavirus pandemic, governments worldwide have enforced strict containment measures such as lockdowns, border closures, quarantining of infected individuals and social distancing. As a result, a considerable proportion of the population employed within schools, universities, companies, and government agencies have transitioned to remotely working from home with no signs of that trend slowing. [4]. In a report published by Gartner, 74% of CFO's interviewed intended to permanently move at least 5% of their workforce to work from home to save costs [5] after the pandemic. These changes, however, have ultimately increased the attack surface for cyber criminals [6].

In the past, Cybercriminals have launched cyber-attacks close to significant world events such as the coronavirus pandemic. [20] surveyed 319 people post Hurricane Harvey to determine if cybercriminals use natural disasters as a basis to launch phishing campaigns [20]. Just over 36% of respondents received phishing emails during the disaster, and surprisingly 10% of respondents clicked on the links provided. Similarly, throughout the COVID19 health crisis, cybercriminals have become highly active by taking advantage of the massive increase in workload, education, and leisure-related online activity. Most of the malicious activity has included fraudulent emails and messages via social media, which contain malicious links or attachments. Malicious cyber actors are actively targeting individuals and organisations with COVID-19 related applications, making it difficult for individuals to detect malware [19]. For example, front line health staff at UW Medicine heavily utilised telehealth to assist patients remotely during the COVID-19 pandemic. During this time, staff reported a dramatic increase in phishing emails (spear phishing) enticing them to download malware [6] via malicious links. In addition to hindering front line staff in an already crippled health system, the spread of malware or ransomware on health care systems can delay the diagnoses and treatment of COVID-19 patients.

To tackle this growing concern, this paper proposes a model to detect COVID-19 URL's shortly after domain name registration using machine learning based classification techniques. We have used recent malicious domain names detected during the COVID-19 pandemic and process the data for feature extraction using our developed tools. Our model leverages a minimal number of features which are available at the time of domain registration, such as the number of hyphens, the number of numeric characters, the Shannon's entropy, the URL length and categorised based on the malicious threat weighting. Although entropy has been used in previous literature to improve detection performance, empirical data in this study suggest that entropy provides better performance in our experimental evaluation. The rest of the paper is organised as follows; section 2 outlines current COVID-19 related literature and section 3 describes our proposed model. Section 4 describes the data set and experimental setup. Section 5 discusses the experimental results. Finally, section 6 provides the conclusion and future work followed by references.

2. **Literature review**

During this COVID19 time, the cybercriminals are taking advantage of sending fraudulent email and messages through social media that attempt to trick you into clicking on malicious links or opening attachments. The malicious cyber actors have been using the Coronavirus as a guise to lure and actively target individuals and Australian organizations with COVID-19 related malicious activities, making it difficult for the individual to detect or unknowingly install malware [26]. Alarmingly, since January 2020 more than 160 thousand new malicious domain names have been created containing COVID-19 related keywords [27]. The malicious activities increase by 2000% due to COVID19 [28]. In response to this trend, the Australian governments recently issued notices educating users not to click COVID-19 related links [29]. Whilst this approach may prevent some phishing attacks it does not stop unsuspecting users from clicking malicious links. Unfortunately, the number of phishing attacks have risen as more companies shift their workforce online [30]. This is alarming considering at least 36% of all data breaches reported in Australia before the COVID-19 crisis originated from phishing campaigns [31].

## 2.1 Emerging statistics of Malicious attack during COVID-19 outbreak

In this section, we are illustrating the senarion of growing malicious attack patterns and attack types of malicious software during COVID-19 pandemic.

### 2.1.1. Sample of malware during covid-19

During covid-19 the malware samples increses exponentially. The following Fig.1 illustrates the graphical representation of malware trends from April 2019 to April 2020 [33]. The figure demonstrates a simple comparison of malware samples from pre-covid-19 (April'19 – January'20) and during covid-19 (Feb'20 to April'20). It has been clear that there is a sharp jump in the number of samples posted from Feb'20 to end of March 2020 when the COVID-19 spread all over the world.

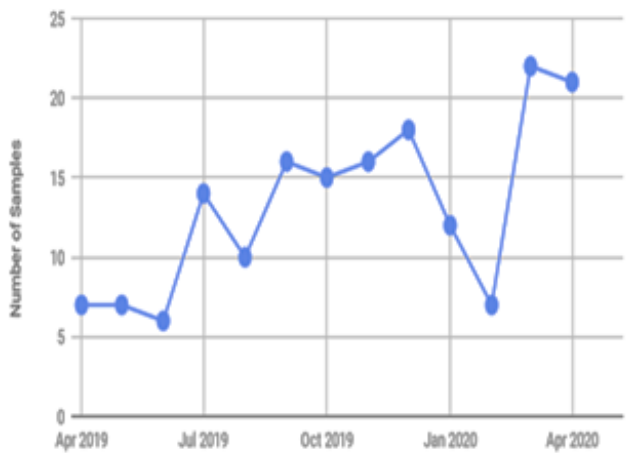

Figure 1: The trend of malware samples posted between April 2019 and April 2020 [33].

Fig. 2 illustrates the number of quick scan files submitted to falcon sandbox [34]. It has been sown that the number of samples gradually increases from February 2020 to end of May 2020, which indicates that approximate ~3000 samples increases within a couple of months during COVID-19 pandemic.

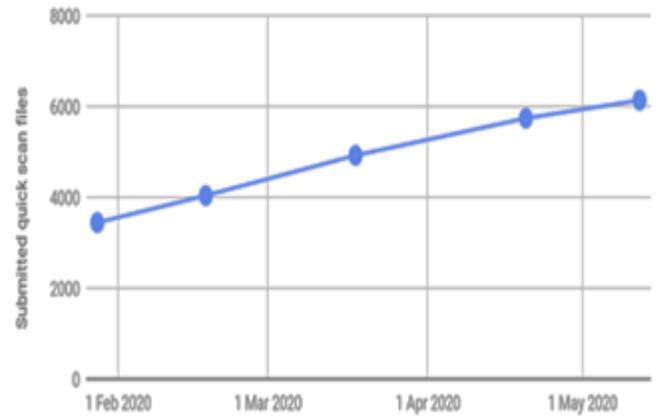

Figure 2: Sample number of quick scan files submitted to Falcon Sandbox. [34]

### 2.1.2 Trend of the attack statistics

This subsection describes the statistics of various attacks in the relation to COVID-19 pandemic. For instance, in Figure 3, we can see the growing number of malware families and its types from falcon sandbox [34]. These are mainly the top 20 families of malware attack statistics between ends of January to end of May 2020.

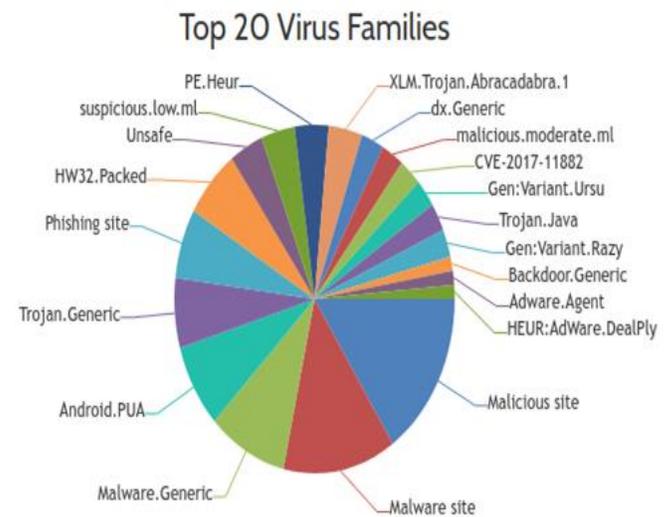

Figure 3: The top 20 virus families between 26 Jan 2020 and 24 May 2020. [34]

The distiribution of malicious file types has been illustrated in Fig. 4 which has been collected from Falcon Sandbox [34]. It is clear that the utmost part of the distribution is malicious URL and then peexe executable, which revealed that during the COVID-19 pandemic a huge number of malicious URL spread through the network traffic.

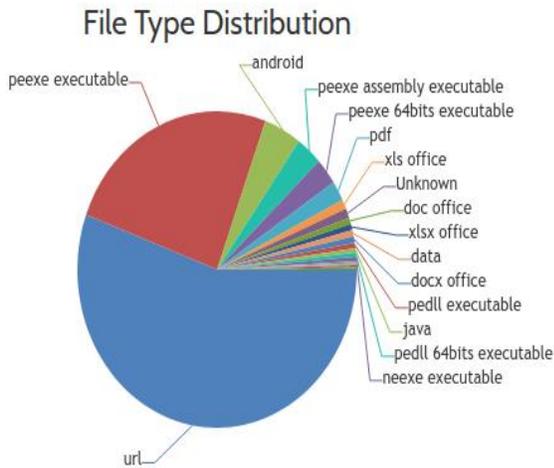

Figure 4. File type distribution between 26 Jan 2020 and 24 May 2020. [34]

### 2.1.3. Phishiing Attack

Phishing has become the most popular attack vector for cybercriminals and its impact of this attack is significant since it can involve the risk of identity theft and financial losses. The phishing scam and spam emails are growing enormously during COVID-19. The scammer is using the government and other potential domains to spread the scam or phishing attack to the user by using COVID related temptations [38]. The user is unknowingly using or following the URLs of instructions and loses their classified data and sometimes compromises the computers or user to make money. The Fig. 5 illustrates a sample of phishing scam which uses Australian government credentials using COVID-19 information [38].

The distribution of COVID-19 related phishing campaigns has become so prevalent and problematic, that government agencies around the world have released public statements to prevent users from clicking on malicious links [18].

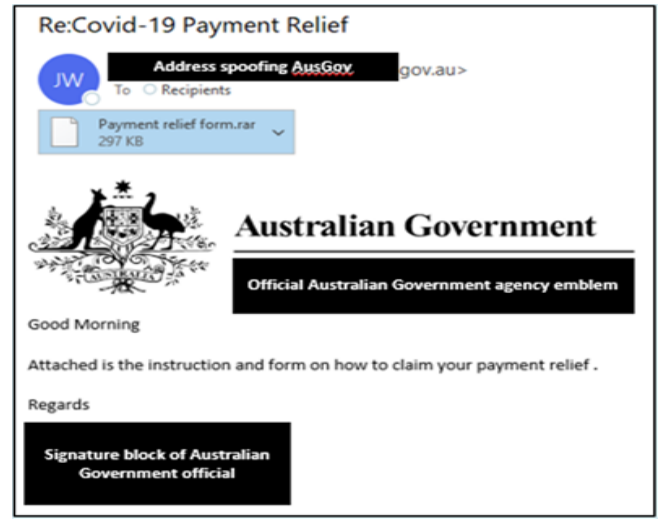

Figure 5: Example email phishing scam pretending to be coming from the Australian Government [38]

Despite best efforts to minimise phishing related attacks, cybercriminals constantly change their tactics to minimise detection and increase the success of phishing campaigns [25]. To be successful, phishing relies heavily on deceiving victims into downloading malware or disclosing personal information. To achieve this, phishers must jump through several hoops before infecting their victim. Even if a malicious email successfully progresses through spam blockers, the content of the text must then convince the recipient to follow the call to action. As a result, phishers increase clickthrough rate by invoking a sense of urgency, concern, or threat [8]. Cybercriminals also commonly use obfuscation to minimise suspicion by mimicking legitimate URL's [4]. This tactic was observed during the pandemic, and alarmingly since January 2020, more than 160 thousand new malicious domain names were registered to contain COVID-19 related keywords [2] for example, "covid19-guidelines.com" [18].

A crucial step in mitigating phishing attacks is to prevent people from clicking the malicious link in the first place. A number of methods have been proposed throughout literature to achieve this, such as user education, URL blacklists and machine learning.

Minimising the impact of phishing via user education has been widely studied and is one of the preventative pillars heavily utilised by government departments. However, ultimately this approach does not prevent phishing from reaching the inbox of unsuspecting users. This is particularly problematic for younger IT users who have very low awareness of phishing or social engineering.

Blacklists are widespread and offer protection against verified phishing sites. Initiatives such as PhishTank, Google Safe Browsing and OpenPhish maintain huge lists of malicious links, however offer little value in detecting new malicious links (otherwise known as zero-hour links)[8, 24] . For a malicious link to be included in a blacklist, it must first be discovered, shared and then validated. This creates a significant window of opportunity which can be exploited by cybercriminals.

There is a significant lag between phishing attacks and the blacklisting of the corresponding site as illustrated in [17], where authors noted that a staggering 47% to 83% of phishing URL's were blacklisted after 12 hours and often drop off the blacklist only to return later. Despite their static properties, the popularity of Blacklists is a result of their ease of implementation and high accuracy. This is largely because they are essentially just a list of malicious links. However, cybercriminals can easily bypass blacklists by obfuscating characters in the path to avoid detection [14]. Overall, the task of managing an updated list of malicious links is an exhausting task and offers little value in preventing zero-hour phishing attacks.

### 2.1.4  Taxonomy of attack patterns

This subsection describes the categorizaotn of threat and attack patterns generated durign COVID-19 outbreak due to remote work and remote access. We have explored various data from Kaspersky from their collections [35-37]. As more users have switched to working from home arrangements, brute force attacks on Remote Desktop Protocol have increased substantially during the first four months of 2020 as shown in Fig. 7 [36].

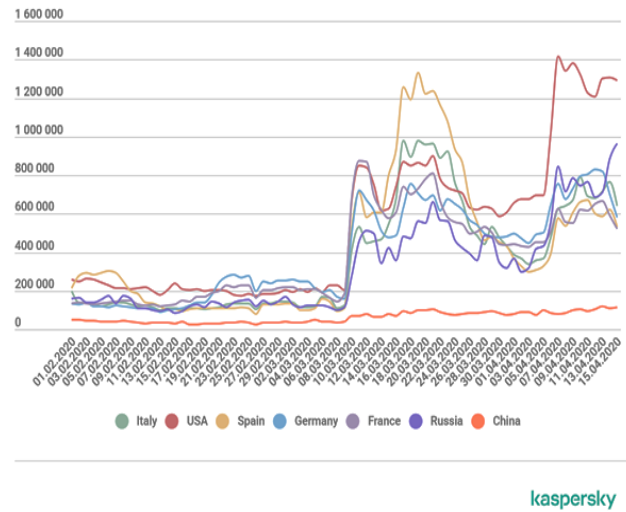

Figure 7: Number of brute force attacks against Remote Desktop Protocol systems reported by Kaspersky from Jan 2020 to Apr 2020 [36].

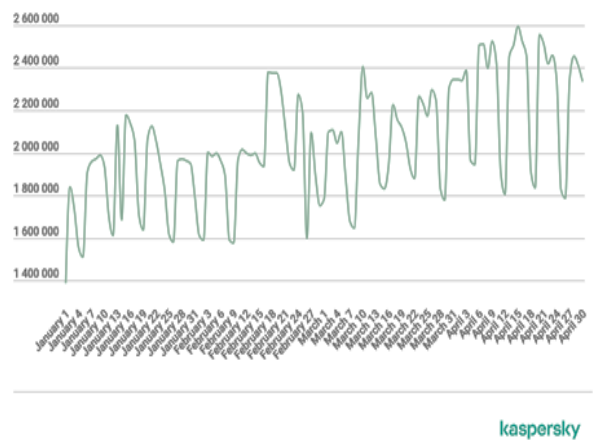

Figure 8: Number of web-based attacks reported by Kaspersky from Jan 2020 to Apr 2020 [36].

We can see from Fig. 7 that most of the growing attack in between March 2020 to mid-April. The attack vector has a sharp increase in early March 2020 and most of the attacks in the region where COVID has impacted more during that time period. There are also the increased demand for remote entertainment, web-based attacks. The Fig. 8 illustrates the web-based attack, from January 2020 to April 2020 of remote entertainment and web contents, have increased by 25% in the first four months of 2020 [36].

Fig. 9 illustrates the DoS attack statistics reported by the first quarter of 2020 [37], which indicates that some of the DoS attacks are high during February 2020 when COVID -19 is spreading around the world.

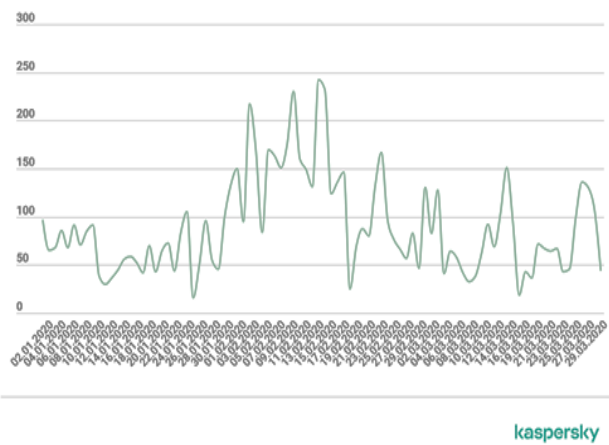

Figure 9: DDoS attacks reported by Kaspersky during Q1 2020 [37].

Fig. 10 demonstrates the geographical attack distribution reported by the first quarter of 2020 [37], which indicates that attack region is high in USA followed in Netherland, Germany and France. It is obvious that those regions were impacted havily from COVID-19, during the first quarter of 2020. It is clear that the maliciour intruder targeted the countries during pandemic when COVID -19 is spreading around the world. Kaspersky also demonstrated the attack pattern that around 40% attack origins from the USA which are most vulnerable in COVID-19 [37].

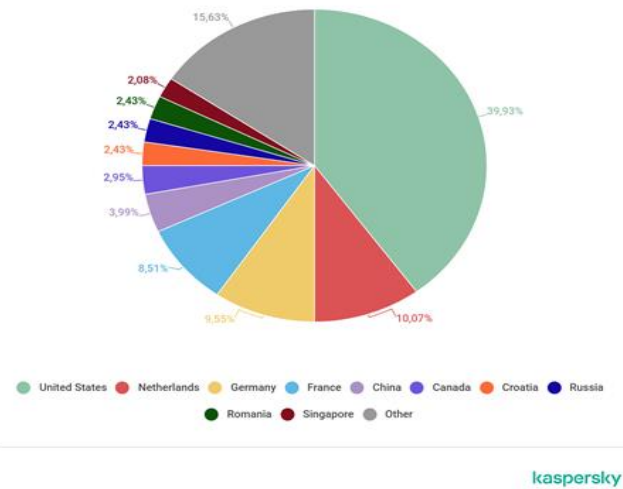

Figure 10: Geographical attack distribution reported by Kaspersky during Q1 2020 [37].

### 2.2. Detecting malicious links with machine learning

The use of machine learning to detect malicious links has been studied extensively throughout literature. The bulk of these studies focus on the detection of malicious links using features such as the code of the corresponding page, the content of the website, the infrastructure details underpinning the UR or the string of characters within the URL (lexical features),. The detection of malicious links via the lexical features within the URL has been shown to be fast and is low risk since it does not require navigation into the malicious link [15].

Studies such as [1-10] focus on the detection of malicious links using lexical features. [24], compared the performance of different machine learning algorithms using lexical features and distance calculations. Although the authors use only the subdomain, domain name and TLD to extract a small number of features, they attain an accuracy rate of 95% using SVM. By comparing the performance of different lexical features, they show a correlation between the domain name length, the number of hyphens and number of numeric characters within malicious links. Other studies such as [1] achieve impressive detection accuracies, however non-lexical features such as infrastructure information and page rank contribute to the final detection accuracy.[39] take a different approach and use natural language processing techniques to detect phishing URL's. In their study, the authors extract brand name similarity, word randomness and over 40 natural language processing (NLP) features from a dataset of over 37000 phishing URL's taken from PhishTank. Although the authors obtain an accuracy rating of 97% using RandomForrest, their approach requires a significant amount of pre-processing.

Other studies such as [13][21] also use word entropy to detect malicious links. [13] use Shannon's Entropy in parallel with other lexical features and achieve an impressive 99% accuracy. [21] separate URL's into n-grams and prove the effectiveness of this approach using Shannon's entropy. The authors make a positive contribution by proving that character distributions within phishing URL's are skewed due to obfuscation techniques used by Phishers. Taking this finding into account, we also include Shannon's entropy as a feature throughout our study.

Overall, most studies on the detection of malicious links using the lexical features within a URL obtain their datasets from known blacklist sites such as PhishTank. This approach is accurate since most of these URL's have been verified to be malicious, however, the bulk of these URL's are harvested late in the attack lifecycle. For this reason, many studies improve the detection accuracy of their models using features not available at the time of domain name registration, such as special characters. To minimise the risk of malicious links, it is ideal to detect malicious domain names shortly after registration before they are circulated as malicious links. As [21] shown in their study, many phishing links obfuscate characters in domain names, indicating possible malicious intention at the time of domain name registration. Therefore, the aim of this research is to:

The aim of this research is to:

1. detect malicious links early in the attack lifecycle, shortly after domain name registration and before they are circulated as malicious links
2. Combat malicious URL's related to COVID-19, and their associated security risks posed by COVID-19 online fraud campaigns.
3. Identify effective measures to mitigate COVID-19 related threats, and safeguard the confidentiality, integrity and availability of data held at the individual and organisational level.

## 3. Proposed detection model

To combat the influx of malicious URL's related to the coronavirus, we propose a mode which detects malicious URL's related to COVID-19. The detection of malicious URL's via the lexical features present in the hostname is fast and low risk since navigation into the domain name is required. Most detection models throughout literature have been designed to detect URL's from popular blacklisting sites such as PhishTank. However, most of the features utilised in these models are not available at the time of registration. For example, characters such as percent (%), curly brackets ({}), and the equal sign (=) are commonly used by cybercriminals to obfuscate phishing URL's but cannot be used whilst registering a domain.

Our proposed model can detect malicious COVID-19 URL's shortly after registration which is early in the attack lifecycle. The following section describes our proposed model to classify malicious URLs. Fig. 11 illustrates the framework of our proposed detection model which is comprised of three main components: -

## 3.1 Data pre-processing

**Comparison engine**

Our study focuses on the detection of COVID-19 related URL's. Therefore, the model searches for COVID-19 related URL's registered for malicious intent. Initially, URL's (both malicious and benign) are ingested into the comparison engine which searches for COVID-19 related keywords and obfuscated variants such as "C0vid", "Cov1d" and "C0ronavirus".

**Dataset processing**

After locating COVID-19 related domain names, all inbound URL's must be standardized for feature selection. The scope of this study focuses on the detection of domain names registered for malicious intent. Therefore, malicious links containing characters which are not available at the time of domain name registration are stripped away. URL components present such as the protocols, ports, paths, and query parameters are removed, leaving only the domain name, top-level domain (TLD) and the second-level domain (SLD).

URL Components removed during processing:

1) **Protocol** – Application layer protocol used to access the internet services. For example, http, https, ftp, etc.
2) **Ports** – The communication port used to access the service. For example, 80, 443, etc.
3) **Path** – The destination of the file on the target server. For example, path/file.php
4) **Query** – Request parameters forwarded to the target server. For example, userId=01.

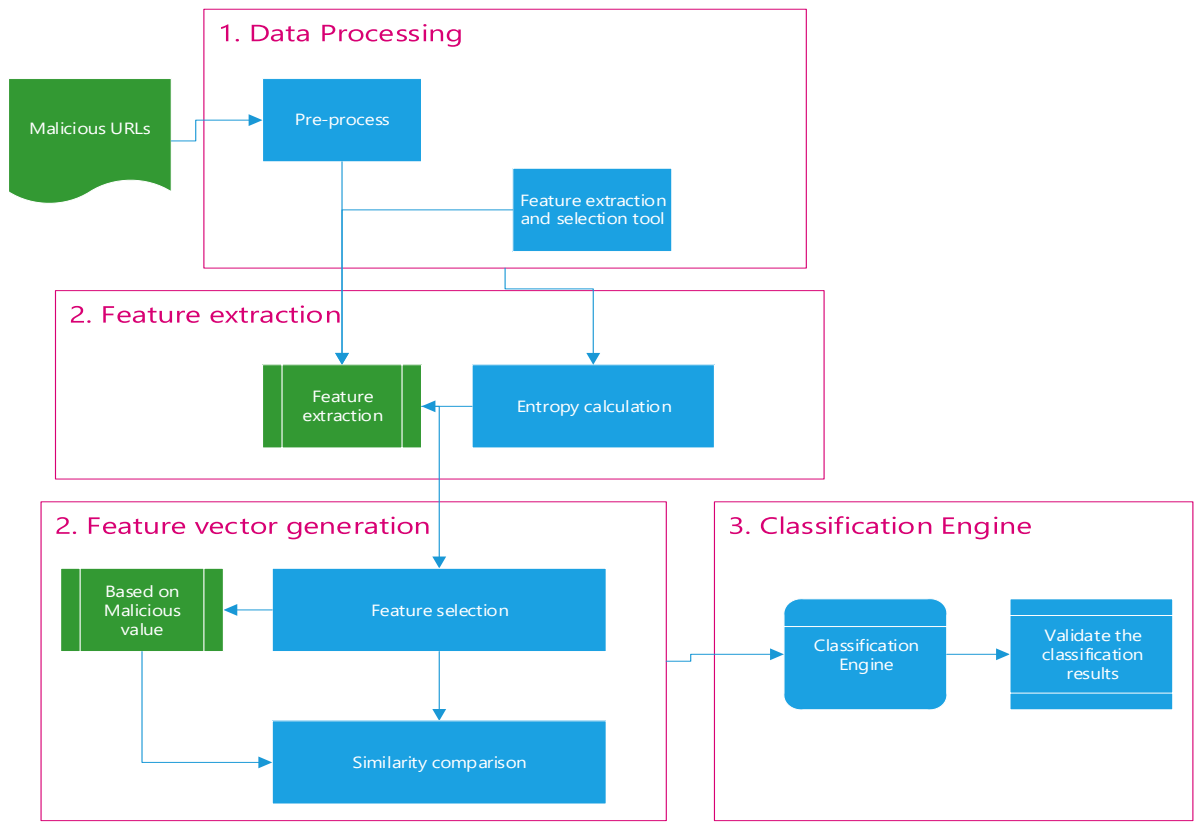

Figure 11. Proposed detection model

URL Components retained and used for feature selection:

a) **Domain name** – The registered identification string. For example, google, apple, amazon, etc.
b) **Top Level Domain (TLD)** – Domains at the highest level of the domain name system (DNS). For example, .com, .edu, .net etc.
c) **Second Level Domain (SLD)** – Domains directly below the TLD, for example .co, .au, .nz, etc.

### 3.2 Feature selection

**Lexical features**

Lexical features were successfully used in [3][11][13][23][24][40][41] to detect malicious URL's. The only characters permissible at the time of domain name registration include the 26 letters of the alphabet (A-Z), numbers (0-9) and hyphens except at the start and end of the domain name string. Although, previous studies included special characters such as percent (%), curly brackets ({}) or hash (#) as part of their features, these are not available at the time of registration. Therefore, special characters have not been used throughout this study to detect malicious links. The analysis of the lexical features within a dataset comprised of malicious links illustrated in Fig. 12 where the x-axis presents the number of URL's and y-axis presents the feature set.

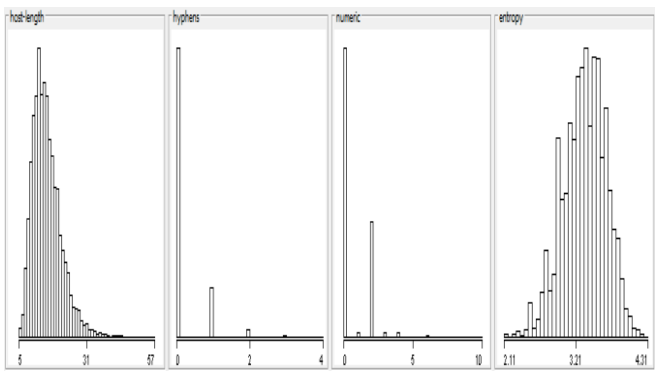

Figure 12: Analysis of the lexical features within a dataset comprised of malicious links.

On the other hand, the Fig. 13 demonstrated the analysis of the lexical features within a dataset comprised of legitimate links, where the x-axis presents the number of URL's and the y-axis presents the feature set.

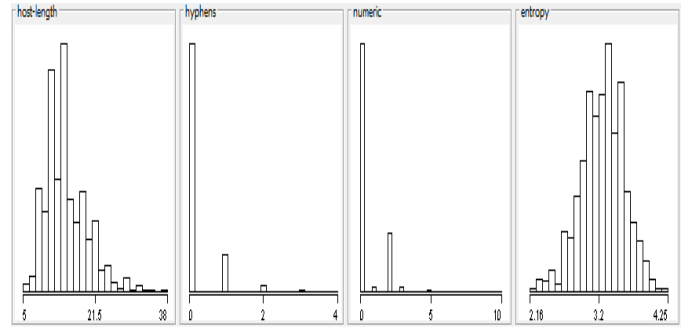

Figure 13: Analysis of the lexical features within a dataset comprised of legitimate links.

To fit with this criteria, we take a similar approach to [13][24][40] who successfully detected malicious links using the number of hyphens and numbers within the domain name string. The features selected to train our model are shown in Table 1.

### 3.3 Feature extraction

Our model uses a minimum number of features. In total, five features have been used throughout this study. These include the length of the domain name, the count of hyphens, the count of numerical characters, entropy calculation and the risk rating.

a.) **Length of domain name** – The total number of characters in the domain name are included as a feature. All other components such as TLD and SLD are not factored into the length calculation.
b.) **Count of Hyphens** – The total number of hyphens present within the domain name. For example, www.example-website.com contains one hyphen character.
c.) **Count of numerical characters** – The total number of numeric characters (from 0 – 9) present in the domain name. For example, examp1e-webs1te.com, contains 2 numeric characters.

## 3.4 Entropy calculation

In addition to the before mentioned lexical features, the Shannon's entropy calculation of each domain name is calculated. Cyber criminals often use obfuscation to confuse and lure victims by mimicking legitimate URL's or masking suspicious ones [4]. Therefore, the randomness factor of each URL was included in this study [13][21] showed that malicious URL's have higher entropy calculations on average when compared to legitimate URL's. Shannon's entropy is calculated using the following equation:

$$H(x) = -\sum_{i=0}^{n} p(x_i) \log_b p(x_i)$$

b.) malicious URL's which have been previously flagged by blacklists.

Where H(x) determines the Shannon entropy and x is the string being assessed. A higher value of H(x) indicates more randomness in string x.

**Malicious value**

The malicious value determines if a URL has been registered for malicious intent. The model takes known URL's which are known to be both malicious and benign and delegates a risk rating based off the likelihood that the corresponding URL is malicious. Each URL is binary classed as follows:

a.) **Malicious** - High probability the URL is used for malicious intent such as harvesting credentials or contains malicious code. The corresponding URL's resemble known c.) **Benign** – Low probability that the URL is malicious. Does not resemble other known malicious URL's.

Table 1: List of lexical features used in this study

| Number | Feature | Description | Previous studies using this feature |
|---|---|---|---|
| 1 | Length | Length of the domain name | [3][13][23][24][40] |
| 2 | - | Number of hyphens within the domain name | [13][24][40] |
| 3 | [0-9] | Number of numeric characters within the domain name | ; [13][24][39] |
| 4 | Entropy | Shannon's entropy calculation of the domain name | [13][21] |
| 5 | Malicious value | Binary rating. 1 for malicious URL's and 0 for benign URL's. | N/A |

## 4. Classification engine

The classification engine is tested and trained with WEKA's implementation of the following classification algorithms; Support Vector Machines, kNN, Naïve Bayes, Logistic regression and AdaBoostM1.

# 4. Data collection and Experimental setup

This section describes the collection of malicious data and our proposed experimental setup using ML techniques.

**4.1 Data Colelction**

To determine which domain names are a potential cyber threat, two classes of data were required: benign and malicious domains. Malicious domains were obtained from DomainTools and was comprised of newly registered domain names related to the coronavirus [2]. DomainTools provides threat intelligence of new and discovered domains and has been used in previous studies [42]All domain names in the dataset contain a corresponding risk rating ranging from 70 to 100 which is a strong indicator of an existing or impending threat. To determine a malicious threat rating, DomainTools measure the proximity to known malicious sites using the domain name, registration information and the underlying infrastructure details.

To gather a list of benign domain names, a publicly available list of recently registered domain names was extracted from WhoisDS from the 7th of April until the 25th of April 2020 [9].The dataset was then filtered for coronavirus related keywords such as "Covid-19", "cov-19", "coronavirus" and "carronavirus". During this time, a total of 27,841 domain names were registered across the internet containing coronavirus related keywords.

After filtering, each domain name was data-matched with the DomainTools dataset which contained 154,292 malicious domain names related to the coronavirus. Any domain name not available within both datasets was assumed to be benign. Post processing, only 1,573 new domain names were determined to be benign. A further 6,321 malicious domain names were added to the benign domains to create a balanced ratio 20:80. The following Table 2 shows the experimental data set.

Table 2: Experimental Data set.

| Dataset | Number of URL's used |
|---|---|
| WhoisDS | 1,573 legitimate domains related to COVID-19 |
| DomainTools | 6,321 malicious domains related to COVID-19 (randomly selected after matching with the WhoisDS dataset) |

**4.2 Analysis of lexical features**

We performed an analysis of both the benign and malicious links in the dataset. The mean of the host length, number of hyphens, number of numeric characters and the entropy of malicious links were all higher than legitimate links, as shown in the Table 3. This validates previous studies such as [24] who observed a positive correlation between the host length and the number of numeric characters within malicious links. Entropy was also noted to be higher in [13][21] on average in malicious links. This was also observed with malicious links within this study.

Table 3: Feature attributes.

| Feature | Legitimate links | Malicious links |
|---|---|---|
| Host length | 14.851 | 16.024 |
| Hyphens | 0.19 | 0.2 |
| Numeric Characters | 0.455 | 0.647 |
| Entropy | 3.262 | 3.342 |

**4.2 Feature extraction**

Python and Pandas were used to extract the selected features from the URL's. The calculation of each URL's Shannon's entropy value is shown in algorithm 1. Each domain is split at the Top-Level Domain (TLD) using a dot (.). This exposes the domain name to extract the selected features. Using Python, we calculate the number of characters in the domain name, the number of hyphens and the number of numeric characters.

**Algorithm 1.**

```
def entropy(s):
    p, lns = Counter(s), float(len(s))
    return -sum( count/lns * math.log(count/lns, 2) for count in p.values())
```

### 4.3 Classification technique

In our classification process, we input the generated feature vectors into the WEKA classification system for which we have written an interface. In all experiments, 10-fold cross validation is applied to ensure a thorough mixing of features. In this procedure, we first select one group of malicious data from a particular data set and divide it into ten portions of equal size; then we select cleanware data of the same size as the group of malware data and also divide it into ten portions. The portions are then tested against each other as demonstrated in Figure 14.

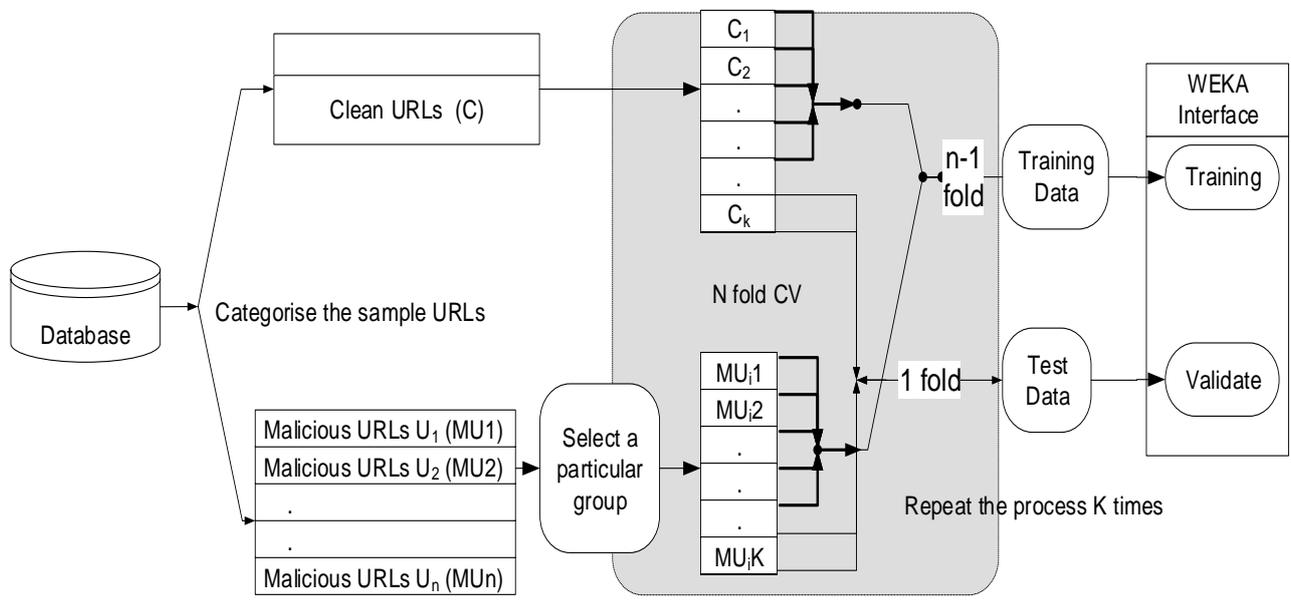

Figure 14: Classificaiton using n-fold technique

To establish the training set, our detection engine takes nine portions from each of the malware and cleanware to set up the training set. The remaining portions from both malware and cleanware are used for the testing set. As is customary, the training set is used to establish the model and the testing set is used to validate it. The whole process is repeated so that every portion of both malware and cleanware is chosen as testing data; the results are then averaged. To ensure that the input vectors are trained and tested over a broad spectrum of classifiers, we chose the classifier/s from WEKA as they represent differing approaches to statistical analysis of data.

## 5. Experimental results

This section describes the classification results based on the dataset we have prepared after entropy calculation. In our experiment we have used WEKA in our all experiment.

Table 4 to Table 8 respresents the classification results using various ML based classification techniques. Its has been shows that the kNN , NB ( Naïve Bayes) and Logistic regration provide better performance without entropy , however SVM and Adaboost has demonstrated no difference between entropy calculation. In our future work we are investigating it with modifying various paramentes and see why entropy calculation not impacting in classification results.

Table 4: Classification results from SVM

|  | TP | FP | Precision | Recall | ROC |
|---|---|---|---|---|---|
| SVM (with entropy) | 0.978 | 0.006 | 0.979 | 0.978 | 0.995 |
| SVM (without entropy) | 0.978 | 0.006 | 0.979 | 0.978 | 0.995 |

Table 5 : Classification results using kNN

|  | TP | FP | Precision | Recall | ROC |
|---|---|---|---|---|---|
| kNN (with entropy) | 0.970 | 0.010 | 0.970 | 0.977 | 0.987 |
| **kNN (without entropy)** | **0.992** | **0.002** | **0.992** | **0.992** | **0.999** |

Table 6 : Classification results using Naïve Bayes

|  | TP | FP | Precision | Recall | ROC |
|---|---|---|---|---|---|
| Naive Bayes (with entropy) | 0.963 | 0.011 | 0.966 | 0.963 | 0.998 |
| Naive Bayes (without entropy) | 0.966 | 0.010 | 0.968 | 0.966 | 0.998 |

Table 7 : Classification results using Logistic Regression

|  | TP | FP | Precision | Recall | ROC |
|---|---|---|---|---|---|
| Logistic Regression (with entropy) | 0.999 | 0 | 0.999 | 0.999 | 1.0 |
| Logistic Regression (without entropy) | 1.0 | 1.0 | 1.0 | 1.0 | 1.0 |

Table 8 : Classification results using AdaBoostM1

|  | TP | FP | Precision | Recall | ROC |
|---|---|---|---|---|---|
| AdaBoostM1 (with entropy) | 0.652 | 0.116 | N/A | 0.652 | 0.849 |
| AdaBoostM1 (without entropy) | 0.652 | 0.116 | N/A | 0.652 | 0.849 |

**Comparison of the accuracy of classifiers:**

WEKA V3.8.4 was used to finalise our results which can be found in Table 6. The algorithms used were based off WEKA's implementation of the classification models: Support Vector Machines, kNN, Naïve Bayes, Logistic regression and AdaBoostM1. WEKA's default settings were maintained throughout the study. Each algorithm was run twice (with and without the entropy calculations) to measure the significance of the entropy calculations towards the final result.

Table 9 : Comparrison of classification results

|  | TP | FP | Precision | Recall | ROC |
|---|---|---|---|---|---|
| SVM (with entropy) | 0.978 | 0.006 | 0.979 | 0.978 | 0.995 |
| SVM (without entropy) | 0.978 | 0.006 | 0.979 | 0.978 | 0.995 |
| kNN (with entropy) | 0.970 | 0.010 | 0.970 | 0.977 | 0.987 |
| **kNN (without entropy)** | **0.992** | **0.002** | **0.992** | **0.992** | **0.999** |
| Naive Bayes (with entropy) | 0.963 | 0.011 | 0.966 | 0.963 | 0.998 |
| Naive Bayes (without entropy) | 0.966 | 0.010 | 0.968 | 0.966 | 0.998 |
| Logistic Regression (with entropy) | 0.999 | 0 | 0.999 | 0.999 | 1.0 |
| Logistic Regression (without entropy) | 1.0 | 1.0 | 1.0 | 1.0 | 1.0 |
| AdaBoostM1 (with entropy) | 0.652 | 0.116 | N/A | 0.652 | 0.849 |
| AdaBoostM1 (without entropy) | 0.652 | 0.116 | N/A | 0.652 | 0.849 |

Logistic regression produced unreliable results and hence were discarded. Using kNN without the entropy calculation in the dataset produced the highest accuracy (99.2%) and the lowest false positive rate (2%). The second highest result was produced when using SVM with and without the entropy calculation. Overall, the entropy calculation offered little value when detecting malicious links. When kNN was run with the entropy calculation the overall accuracy decreased to 97% and false positive rate (FPR) increased to 10%. No difference in accuracy or FPR was observed when using SVM.

## 6. Conclusion and Future work

In this paper, we proposed a framework to detect malicious domain names containing COVID related keywords. Using only 5 lexical features, our model detected malicious domain names with a 99.2% accuracy rate. To achieve this, we trained and tested our model using 7849 domain names from WhoisDS and DomainTools. Although the entropy of malicious domains was higher on average than benign domain names, entropy added little value to the overall accuracy rate. Infact, the best accuracy rate was achieved using kNN without the use of entropy.

Overall, our model offers a promising solution to minimising COVID related phishing and malware attacks by detecting malicious domain names, early in the attack lifecycle. This is due to its ability to detect malicious URL's with a high accuracy using only the domain name and a minimal number of features.

Future work will investigate the incongruence of entropy as a feature. Unlike other studies which used entropy to improve their detection accuracy, we found that the inclusion of entropy offerered little value. This may be due to the changing tactics of cybercriminals who deliberately register domain names with a lower entropy to evade detection. Thus understanding this phenomena would offer a positive contribution to knowledge.